\documentclass[letterpaper, 10 pt, conference]{ieeeconf}  

\IEEEoverridecommandlockouts                              

\usepackage{graphicx}
\usepackage{epstopdf}          
\usepackage{amsmath} 
\usepackage{amssymb}  
\usepackage{color}

\title{\LARGE \bf
Model Predictive Control for Automotive Climate Control Systems via Value Function Approximation
}

\author{Dennis Kibalama$^{1}$, Yuxing Liu, Stephanie Stockar, Marcello Canova$^{2}$
\thanks{$^{1}$ The Author is with the Department of Electrical Engineering and the Center for Automotive Research (CAR), The Ohio State University, $^{2}$ The Authors are with the Department of Mechanical and Aerospace Engineering and the Center for Automotive Research (CAR), The Ohio State University,
        Columbus, OH 43212, USA
        {\tt\small kibalama.3@osu.edu}}%
}

\begin{document}

\maketitle
\thispagestyle{empty}
\pagestyle{empty}

\begin{abstract}
Among the auxiliary loads in light-duty vehicles, the air conditioning system is the single largest energy consumer. For electrified vehicles, the impact of heating and cooling loads becomes even more significant, as they compete with the powertrain for battery energy use and can significantly reduce the range or performance. While considerable work has been made in the field of optimal energy management for electrified vehicles and optimization of vehicle velocity for eco-driving, few contributions have addressed the application of energy-optimal control for heating and cooling loads.

This paper proposes an energy management strategy for the thermal management system of an electrified powertrain, based on Model Predictive Control. Starting from a nonlinear model of the vapor compression refrigeration system that captures the dynamics of the refrigerant in the heat exchangers and the power consumption of the system, a constrained multi-objective optimal control problem is formulated to reduce energy consumption while tracking a desired thermal set point. An efficient implementation of MPC is proposed for real-time applications by introducing a terminal cost obtained from the approximation of the global optimal solution.
\end{abstract}

\section{INTRODUCTION}

Heating and cooling loads, such as the air conditioning (A/C) system of the cabin, represent the largest energy consumer among the auxiliary loads in light-duty vehicles \cite{rugh2008vehicle}. The impact of such loads becomes even more significant in energy-efficient vehicles, such as Battery Electric Vehicles (BEVs), or Hybrid Electric Vehicles (HEVs). These powertrains are characterized by a much higher efficient power conversion, hence the auxiliary loads impact the energy consumption and the range more prominently than compared to conventional powertrain that are based on internal combustion engines. In the case of BEVs, where a heat pump system is often used to satisfy the heating or cooling demand of the battery pack, electric machines and power electronics, the auxiliary loads compete with traction power and can heavily drain the battery in extreme weather conditions \cite{chowdhury2018total}. Recent studies have shown that thermal loads can reduce EV range and HEV fuel economy by nearly 40$\%$, depending on the size of the air-conditioner and the driving cycle  \cite{farrington2000impact}.

Several methods have been proposed to reduce the energy use of vehicle climate control systems for xEV applications, including cabin pre-conditioning \cite{kambly2015geographical, jeffers2016climate}, or zonal climate control \cite{jeffers2016climate, zhang2018electric}. On the other hand, the use of energy optimization methods for the control of thermal systems, especially for electrified vehicles, has only been sparsely explored in research. In \cite{zhang2015energy}, the problem of reducing energy consumption of a vehicle air conditioning system while maintaining cabin comfort requirements is formulated as an optimal control problem and solved off-line via Dynamic Programming (DP), to investigate the trade-off among multiple control objectives. The Pontryagin’s Minimum Principle (PMP) is then applied to obtain an online implementable control strategy. Similarly, PMP was employed in \cite{bauer2014thermal} to optimize the thermal management dynamics of a BEV. Model Predictive Control (MPC) has recently been applied to the energy-optimal control of automotive air conditioning systems \cite{huang2017modelling, schaut2019thermal}, showing promising results.

Recent contributions have been made in the field of energy-efficient driving strategies (Eco-Driving), for example, via optimization of HEV energy management and vehicle velocity trajectories, showing considerable potential for energy savings \cite{vahidi2018energy}. In this context, predictive optimization of vehicle auxiliary loads has been considered in the context of Connected and Automated Vehicles (CAVs). For instance, \cite{wang2018model}utilizes vehicle speed preview to coordinate the cabin temperature accordingly, showing that the energy efficiency of the A/C system can be improved by up to 9$\%$.

One of the known issues of MPC is the computation time. In the case of A/C systems, the dynamics of the vehicle velocity and cabin temperature are typically slow compared to the dynamics of the refrigerant in the vapor-compression system, which forces one to adopt long prediction horizons as well as linearize or simplify the plant dynamics \cite{schaut2019thermal}. Furthermore, the equations describing the refrigerant dynamics are typically nonlinear \cite{zhang2013lumped}, which can add further computation requirements and prevent the controller from being implementable in automotive control units.

This paper proposes an energy management strategy for the thermal management system of an electrified powertrain, based on MPC. Starting from an energy-based model of an A/C system that captures the nonlinear dynamics of the refrigerant and the power consumption of the actuators, a constrained multi-objective optimal control problem is formulated to minimize the energy use, while tracking a desired thermal set point. A novel implementation method is proposed to drastically reduce the computation time, allowing online implementation. The concept hinges on defining a terminal cost approximation from a global optimal solution obtained via DP. The approximated terminal cost leads to a sub-optimal solution that achieves very close performance to DP on different drive cycles and thermal load requirements, even when the prediction horizon is reduced to a single step.

\section{OVERVIEW OF THE A/C SYSTEM MODEL}

\begin{table*}[!h]
\begin{equation}\label{E:Energy_AC_10}
\begin{aligned}
  &V_e\left[ \left(1- \bar{\gamma_e}\right) \frac {\partial \left(\rho_l h_l\right)_e}{\partial p_e} + \bar{\gamma_e}  \frac {\partial \left(\rho_g h_g\right)_e}{\partial p_e } + \left(\rho_g h_g -\rho_l h_l\right)_e \frac{\partial \bar{\gamma_e}}{\partial p_e} -1 + \frac{M_{we} c_e}{V_e} \frac{\partial T_e}{\partial p_e}\right] \frac{d p_e}{dt} = \dot Q_{e} + \dot m_r\left(h_{4} - h_{g}\left(p_e\right)\right)\\
  &V_c\left[ \left(1- \bar{\gamma_c}\right) \frac {\partial \left(\rho_l h_l\right)_c}{\partial p_c} + \bar{\gamma_c}  \frac {\partial \left(\rho_g h_g\right)_c}{\partial p_c} + \left(\rho_g h_g -\rho_l h_l\right)_c \frac{\partial \bar{\gamma_c}}{\partial p_c} -1 + \frac{M_{wc} c_c}{V_c} \frac{\partial T_c}{\partial p_c}\right] \frac{d p_c}{dt} = -\dot Q_{c} + \dot m_r\left(h_{2} - h_{l}\left(p_c\right)\right)
\end{aligned}
\end{equation}
\end{table*}

\begin{figure} [!htb]
  \centering
  \includegraphics[width=0.9\columnwidth]{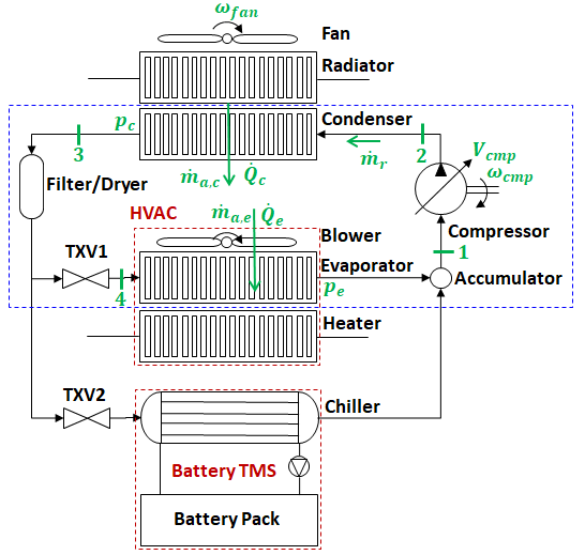}
  \caption{Plant Diagram of the Climate Control System for a HEV.}\label{F:layout}
\end{figure}

The climate control system of a HEV, for instance the one shown in Figure \ref{F:layout}, is based on a simple vapor compression cycle where multiple evaporators in parallel provide the necessary cooling power to the cabin and the battery pack Thermal Management System (TMS) \cite{chowdhury2018total}. In this configuration, the compressor is connected to the engine through a belt, and a variable displacement system is used to control the mass flow rate of refrigerant that circulates through the system and, consequently, the heat transfer rate at the evaporator and condenser. Figure \ref{F:layout} provides a schematic representation of the system, along with the relevant thermodynamic states.

While this study focuses on the energy optimization of the cabin HVAC system, its results can be extended to include the battery TMS, for example by adopting a hierarchical control design approach \cite{amini2020hierarchical}.

Since the A/C system operates in highly transient conditions due to the variability in the compressor speed, cooling demands and vehicle speed (which affects the air flow rate at the condenser), the external inputs that govern the state dynamics may change rapidly and considerably \cite{02_Li2010}. Furthermore, the presence of a phase changing fluid and the different time scales induced by the mass and energy transport pose significant modeling challenges in predicting the state dynamics, namely the pressures in the condenser and evaporator.

This work adopts the lumped-parameter model proposed in \cite{zhang2013lumped}, where the dynamics of the refrigerant in the heat exchangers was obtained by applying mass and energy balance equations, and introducing a set of simplifying assumptions that reduce the model to only two states. The state equations, which predict the pressure in the heat exchangers, are summarized in Eq. \eqref{E:Energy_AC_10}, where $p_e$ and $p_c$ indicate the evaporator and condenser pressure, and the subscripts $2$ and $4$ refer to the thermodynamic conditions marked in Figure \ref{F:layout}. The thermodynamic properties, including density $\rho$, enthalpy $h$, and mean void fraction $\bar{\gamma}$ are mapped from the thermodynamic tables of refrigerant R-1234yf.

To close the model equations, the rate of heat transfer $\dot Q_{e}$ into the evaporator and $\dot Q_{c}$ dissipated by the condenser must be expressed as functions of the thermodynamic states and flow rates. To this end, the $\varepsilon$-NTU method is adopted to define approximated expressions under the assumption that the refrigerant within the two heat exchangers is predominantly two-phase \cite{zhang2013lumped}. The expressions obtained are:
\begin{equation}\label{E:Heat_Transfer}
\begin{aligned}
  \dot Q_{e} &= \dot m_{a,e} c_{p}\left(T_{a,in,e} - T_{e}\left(p_e\right)\right)\cdot \left(1-exp\left(-k_{e}NTU_e\right)\right) \\
  \dot Q_{c} &= \dot m_{a,c} c_{p}\left(T_{c}\left(p_c\right) - T_{a,in,c}\right)\cdot \left(1-exp\left(-k_{c}NTU_c\right)\right) \\
\end{aligned}
\end{equation}
where $T_{e}$ and $T_{c}$ are the saturation temperatures of the refrigerant, which are dependent on the corresponding pressures. The NTU parameter depends on the geometry of the heat exchangers and on the convection heat transfer coefficient computed on the air side \cite{Incropera2011, Ramesh2003}:
\begin{equation}\label{E_NTU}
  NTU_j = \frac {\epsilon_j A_{s_j} [1-F_{fin,j}(1-\eta_{FA,j})]}{{\dot m_{a,j}} c_{p}}, \qquad (j=e,c)
\end{equation}
where $ \epsilon$ is the air heat transfer coefficient, $ A_{S} $ is the heat exchanger external surface area, $F_{fin} $ is the fraction of air-to-structure surface area on fins, $ \eta_{FA} $ is the air side fin efficiency. The numerical values of the above parameters can be retrieved from the heat exchanger specification sheets.

Note that, in Eq. \eqref{E:Heat_Transfer}, the air mass flow rate $\dot m_{a,e}$ at the evaporator depends on the speed of the blower and is considered an external input, while the air flow $\dot m_{a,c}$ at the condenser depends on the vehicle velocity and the radiator fan speed $\omega_{fan}$, which is a control input.

The compressor model computes the refrigerant flow rate $\dot m_r$ and discharge enthalpy as functions of the compressor speed  (external input), the displacement (control input), the thermodynamic conditions at the suction side, and the pressure ratio $PR = p_c / p_e$. Starting from \cite{03_scott2007}, the flow rate $\dot m_r$ and enthalpy $h_2$ are obtained from models of the volumetric efficiency $\eta_v$ and isentropic efficiency $\eta_s$:
\begin{equation}
\label{E:CMP_E}
  \eta_v = \frac {\dot m_r}{V_{cmp} \rho_1 \omega_{cmp}},  \qquad \eta_s = \frac {h_{2s} - h_1}{h_2 - h_1}
\end{equation}
where $\rho_1 = \rho_g\left(p_e\right)$, $h_1 = h_{g}\left(p_e\right)$ are the thermodynamic conditions at the suction side of the compressor, and $h_{2s}$ the ideal (isentropic) enthalpy corresponding to the compression $1\rightarrow2$, which is a function of the pressure ratio $PR$.

The volumetric and isentropic efficiency are modeled as empirical functions of the compressor speed and the pressure ratio \cite{03_scott2007}:
\begin{equation}\label{E:CMP_V}
	\begin{aligned}
		\eta_v &= \frac{a_0 – (a_1+a_2 Z)(PR^{1/\kappa}-1)}{1+a_3 Z^2}\\
		\eta_s &= \frac{{(PR-1)}^{b_0}}{b_1 + b_2 Z + b_3{\left( PR-1\right)}^{b_4}}
	\end{aligned}
\end{equation}
where $Z= V_{cmp}^{1/3} \omega_{cmp} / \sqrt{p_c / \rho_1} $ is the Mach index and $\kappa$ the specific heat ratio. The parameters $a_0, \ldots , a_3$, $b_0, \ldots , b_4$ are identified on the compressor maps provided by the manufacturer. Finally, the power consumed by the compressor and the radiator fan are obtained from the states and control inputs:
\begin{equation}\label{E:Pow}
\begin{aligned}
P_c &= \dot m_r\left(h_2 –h_1\right) \\
  P_f &= \frac{1}{\eta_{e}} \frac{P_{max}}{\omega_{max}^3} \omega_{fan}^3
\end{aligned}
\end{equation}
where $\eta_{e}$ is the efficiency of the electric motor powering the fan. A detailed description of the model parameters and process for validation are provided in \cite{zhang2013lumped}.

\section{OPTIMAL CONTROL PROBLEM FORMULATION}
An optimal control problem is formulated for the A/C system to minimize the total power consumed by the actuators, while tracking a desired evaporator pressure subject to the system dynamics described in the previous section, state and input constraints. The general formulation is given by:
\begin{align}
&\underset{u}{\text{min}} & & J(x,u,w)=\int_{t_0}^{t_f} L(x,u,w)\cdot dt\\
& \text{subject to} & & \dot{x} = f(x,w,u),\ x(t_0)=x_0\\
&    								& & x_{min}\leq x(t)\leq x_{max}\\
&    								& & u_{min}\leq u(t)\leq u_{max}
\end{align}
where $x=\begin{bmatrix} p_e,&p_c \end{bmatrix}^T$ is the state vector,  $u=\begin{bmatrix} \omega_{fan},& V_{cmp}\end{bmatrix}^T$ is the input vector and $w = \begin{bmatrix} \omega_{cmp},& \dot m_{a,e},& V_{veh},& T_{a,in,e},& T_{a,in,c}\end{bmatrix}^T$ is the vector of external inputs, which includes the compressor speed, the vehicle speed (affecting the air flow rate at the condenser, along with the fan), the air flow rate and inlet temperature at the evaporator (affecting the cooling load). Finally, $L(x,u,w)$ is the running cost given by a linear combination of the normalized power and tracking error on the evaporator pressure:
\begin{equation}
L = \left[ \alpha \frac{P_{f}(t) + P_{c}(t)}{P_{base}}+\left(1-\alpha \right) \frac{(p_e(t)-p_{e,ref})^2}{\Delta p_{max}^2}\right]
\end{equation}
where $P_{f}$ and $P_{c}$ are given in Eq. \eqref{E:Pow}, $P_{base}$ is the average power consumed by the baseline strategy, $p_{e,ref}$ is the reference pressure given by the AC setting, $\Delta p_{max}$ is the maximum allowable deviation from the reference and $\alpha$ is a weight that can be used to explore the tradeoff between energy saving and tracking performance. Note that the tracking is directly implemented in the cost function, as opposed to using a deadband approach. For the system under consideration, the latter would results in a controller always operating the system at the upper evaporator pressure limit \cite{mahdavi2017model}.

The state dynamic is discretized using the Euler forward method, and the optimization problem above is solved using DP \cite{bertsekas1995dynamic,sundstrom2009generic}. While DP generates an offline solution, namely a policy that requires knowledge of the complete drive cycle, it provides a benchmark for online control algorithms.

\section{MPC IMPLEMENTATION}
To design an online predictive controller starting from a global optimal (offline) solution, there are several techniques that trade-off optimality across a range of cycles with computation time and complexity. For instance, \cite{rostiti2015rule} shows that the DP results can be used for rule extraction with near optimal results. More recent contributions have focused on MPC, due to its ability to account for multiple control inputs and nonlinear plant dynamics \cite{huang2017modelling, schaut2019thermal,wang2018model}.

Using MPC to design a controller for the A/C system, however, poses unique challenges due to the fast dynamics of the refrigerant pressure, compared to the slowly-varying external inputs such as vehicle speed and cabin flow rate or temperature. To achieve robust and near-optimal solutions, the prediction horizon has to be extended significantly \cite{schaut2019thermal}. In such case, the accuracy of the prediction model is crucial and therefore a plant model linearization is not possible. Between the long prediction horizon and the nonlinear optimization problem, the computation time required for the solution is significant making it prohibitive for real time applications.
To overcome this challenge, this paper introduces a modified cost function for the economic MPC formulation. While the running cost is integrated over the prediction horizon, an approximated minimum cost is applied from the end of the prediction horizon to the end of the driving cycle. This approach is based on Bellman's principle of optimality \cite{geering2007optimal,bertsekas1995dynamic,rawlings2009model} and has been proven effective in reducing computation time while achieving near-optimal solutions, for example in HEV energy management problems \cite{borhan2011mpc}.

For any time $t$ and state $x$, the optimal control sequence can be obtained from:
\begin{equation}
\underset{u}{\text{min}} \int_{t}^{t_f} L(x,u,w,\tau) d\tau
\end{equation}
Based on Bellman’s Principle of Optimality, this is equivalent to solving:
\begin{equation}
\label{E:2}
\underset{u}{\text{min}} \left( \int_{t}^{t+t_H} L(x,u,w,\tau) d\tau + \int_{t+t_H}^{t_f} L(x,u,w,\tau) d\tau \right)
\end{equation}
where $t_H$ is the length of the prediction horizon. Based on the definition of cost-to-go, Eq. \eqref{E:2} is equivalent to:
\begin{equation}
\underset{u}{\text{min}} \left( \int_{t}^{t+t_H} L(x,u,w,\tau)d\tau + V(x(t+t_H),t+t_H) \right)
\end{equation}

For real-time implementation, future external disturbances $w(t)$ and the duration of the cycle $t_f$ are unknown. Therefore, an approximation for the cost-to-go must be obtained, leading to a sub-optimal formulation:
\begin{equation}
\underset{u}{\text{min}} \left( \int_{t}^{t+t_H} L(x,u,w,\tau)d\tau + \hat{V}(x(t+t_H)) \right)
\end{equation}
where $\hat{V}$ is an approximated value function, or ''base heuristic'' 
\cite{powell2007approximate}. This formulation originates from the theory of Approximate Dynamic Programming (ADP) and can be integrated with an MPC if $\hat{V}$ is assumed as the terminal cost \cite{rawlings2009model}. This approach allows one to significantly reduce the prediction horizon while retaining the nonlinearity of the plant dynamics, which overall results in a computationally efficient, near-optimal MPC formulation \cite{rawlings2009model}.

\subsection{Approximation of the Value Function}
Several methods can be used to define the approximated value function $\hat{V}$. For example, in \cite{borhan2011mpc,johannesson2008approximate} the cost-to-go is assumed to be linear with respect to the state, and the parameters of this linear approximation are tuned. Alternatively, the value function can be modeled as a Markov Decision Process and determined via Reinforcement Learning \cite{zhu2020energy}.

This paper proposes to generate the value function approximation by analyzing the cost-to-go function obtained by solving the DP over a set of representative drive cycles. For this study, a single SCO3 cycle with a fixed A/C setting (SCO3 Mid) is considered.
\begin{figure} [!htb]
  \centering
  \includegraphics[width=1.0\columnwidth]{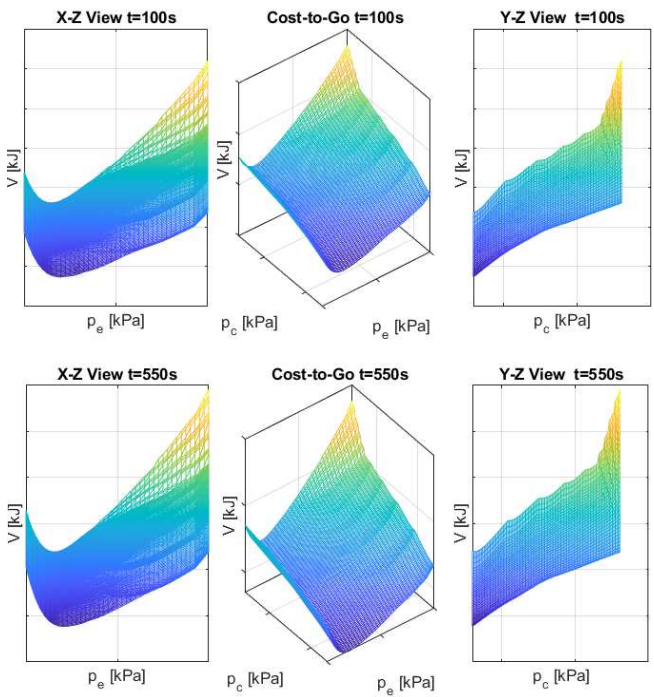}
  \caption{Cost-to-Go Function for the SC03 Cycle ($\alpha=0.1$).}\label{F:C2G}
\end{figure}

The cost-to-go corresponding to the DP solution for the SC03 Mid cycle with $\alpha =0.1$ is shown in Figure \ref{F:C2G}. The value function $V$ decreases with time ($\partial V/\partial t \leq 0$) and retains a similar shape for different time instances. It is also worth noting that a cost of zero is never attained. For a fixed time and condenser pressure $p_c$, the function $V$ is quadratic in $p_e$ and the curvature increases for larger values of $p_c$. Finally, with the exception of high values of $p_c$ and $p_e$, a linear relationship can be observed between $V$ and the condenser pressure.

\begin{figure} [!htb]
  \centering
  \includegraphics[width=1.0\columnwidth]{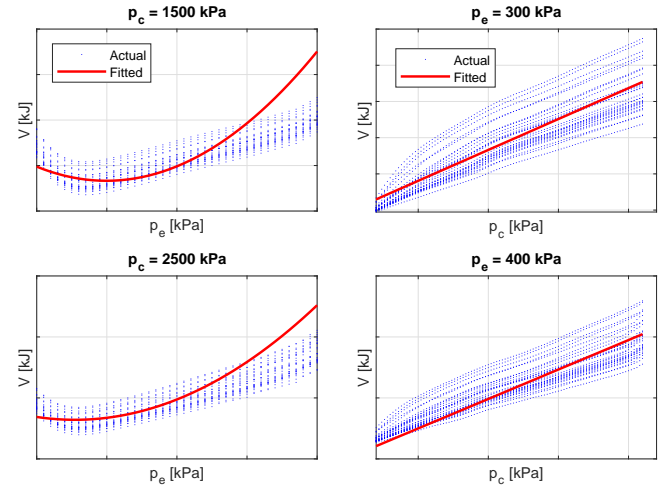}
  \caption{Quadratic Fitting of the Cost-to-Go Function.}\label{F:C2G_fit1}
\end{figure}

To parameterize the reference pressure, the cost-to-go is shifted along the z-axis such that zero cost is attained at $x_r \forall t$:
\begin{equation}
V_{\Delta}(x,t) = V(x,t)-V(x_r,t)
\end{equation}
Finally, after averaging the cost-to-go over time, a quadratic curve fitting is performed on a local region around $x_r$:
\begin{equation}
V_q(x-x_r) = (x-x_r)^TQ(x-x_r)+p^T(x-x_r)
\label{E:Vq}
\end{equation}
The fitting results are shown in Figure \ref{F:C2G_fit1}, with respect to each state variable. The fitting is accurate around the reference, while monotonically increasing as the tracking error becomes larger.

\subsection{MPC Formulation}
Using the above value function approximation as the MPC terminal cost, the receding horizon optimization problem is formulated as:
\begin{align}
&\underset{u}{\text{min}} & & \sum_{k=0}^{N-1} L(x_k,u_k,w_k)+ V_q(x_{N})\\
& \text{subject to} & & x_{k+1} = x_k+f(x,w,u)\cdot \Delta t\\
&    								& & x_{min}\leq x_k\leq x_{max}\\
&    								& & u_{min}\leq u_k\leq u_{max}
\end{align}
where a one-step prediction horizon and a discretization step $\Delta t = 1s$ are selected for this problem. The external inputs $w$ are measured and are kept constant in the receding horizon. The one-step MPC implementation considerably reduces the problem complexity, while still capturing the effect of the external inputs on the state trajectories.

\section{RESULTS}
\begin{figure} [!htb]
  \centering
  \includegraphics[width=1.0\columnwidth]{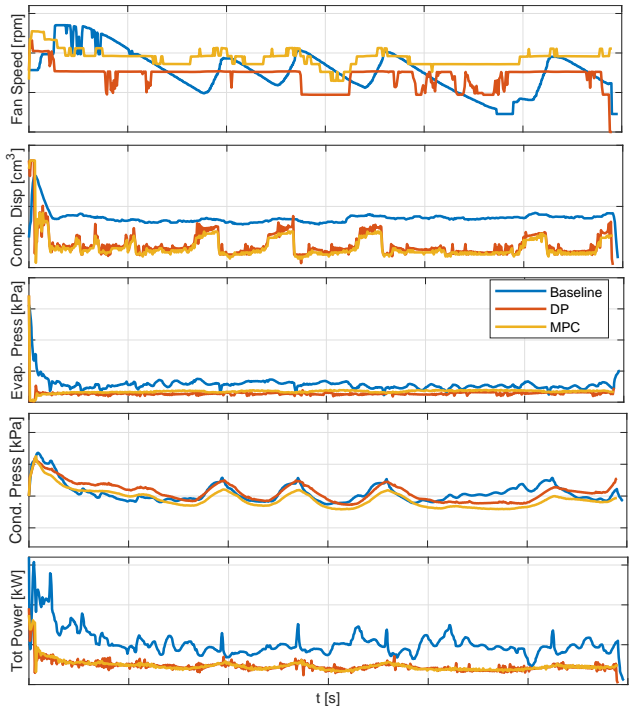}
  \caption{Comparison of MPC against Baseline Solution for the ''SC03 Mid'' Cycle.}\label{F:MPC1}
\end{figure}

\begin{figure} [!htb]
  \centering
  \includegraphics[width=1.0\columnwidth]{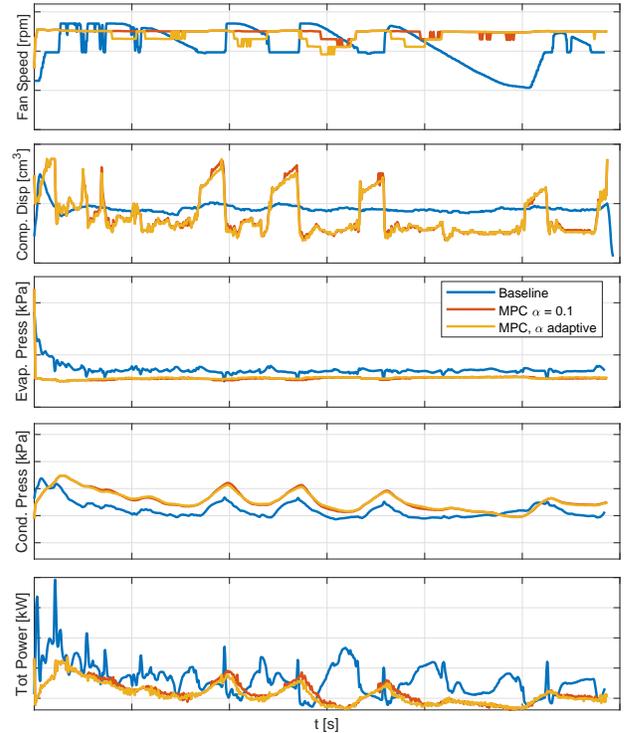}
  \caption{Comparison of MPC against Baseline and DP Solution for ''SC03 High Cycle''.}\label{F:MPC2}
\end{figure}

The MPC defined in the previous section was first tested on the same cycle used to generate the value function approximation. The results are compared against the baseline (production controller) and the global optimal solution obtained from DP. Figure \ref{F:MPC1} shows that, although the fan actuation strategy slightly deviates from DP solution, the power consumption of the one-step MPC solution closely matches the DP solution. Due to the approximation of the value function, the solution of the MPC does not retain the recursive feasibility property that is inherent to the DP solution. However, $V_q$ has been obtained on a cycle that explores a wide enough range of states and control inputs to assume that the system is perturbed around nominal conditions and hence a feasible solution can always be found \cite{hopp1988sensitivity,liu2019recursive}.
The controller with approximated value function is then tested on a different cycle, specifically, the SC03 Low, and the results are shown in Figure \ref{F:MPC2}. The results show that the $\alpha'$ adaptation closely matches the results from implementation with fixed $\alpha'$.

Note that $V_q$ was obtained by approximating the cost-to-go function obtained from DP over one single cycle with a fixed tradeoff weight. However the controller should perform well for a variety of cycles and external conditions. To this end, the weight $\alpha$ in the cost functional $L$ is used as tuning parameter for the online implementation. The approach suits this problem due to the strong continuity in the cost function, which implies that the value function is continuous and therefore the approach is robust against disturbances, such as variations in the A/C settings \cite{tan2011sensitivity}. The tradeoff weight in the MPC is defined as $\alpha'$ to distinguish between the value used in the DP:

	\begin{equation}
		L = \left[ \alpha' \frac{P_{f}(t) + P_{c}(t)}{P_{base}}+\left(1-\alpha' \right) \frac{(p_e(t)-p_{e,ref})^2}{\Delta p_{max}^2}\right] + V_q(x)
	\end{equation}

A comparison between the baseline, DP and the tuned MPC is shown in Figure \ref{F:MPC2}, and a summary of the MPC performance for two different test conditions is given in Table \ref{T:MPC}.

\begin{table}[h]
\caption{Summary of the MPC Performance Metrics}\label{T:MPC}
\begin{center}

\begin{tabular}{|c|c|c|c|c|}
	\hline
	Cycle & Strategy & Total 				& Energy				& Tracking \\
	& 				 & Energy [kJ]  & Savings [$\%$]	& Error [$\%$] \\
	\hline
	SC03 Low	&	Baseline	& 1830	&	N/A	&	N/A	\\
	SC03 Low	&	DP	&	1709	& -6.6$\%$	&	2.08	\\
	SC03 Low	&	MPC, $\alpha'$ adapt	& 1714	&	-6.3$\%$	&	2.5	\\
	\hline
	SC03 Full	&	Baseline	&	1646	&	N/A	&	N/A	\\
	SC03 Full	&	DP	&	1386		&	-15.8$\%$	&	1.08	\\
	SC03 Full 	&	MPC, $\alpha'$ adapt	&	911	&	-23.7$\%$	&	2.0	\\
	\hline
\end{tabular}
\end{center}
\end{table}

Clearly, the quadratic terminal cost approximation $V_q$ in Eq. \eqref{E:Vq} ensures a flat evaporator temperature trajectory, which implies that the MPC implementation is robust against disturbances in vehicle velocity, engine speed, and air inlet temperatures at the evaporator and condenser. The weight $\alpha'$ in the running cost affects the distance between the evaporator pressure and the reference, and can be tuned to compensate for the fact that $V_q$ has been obtained for only one test condition. The turnaround time for the one-step MPC framework is approximately 5-6ms (over various cycles), as evaluated on a desktop computer (with a core i7 processor). The same algorithm was implemented on Rapid-Prototyping hardware and validated in-vehicle with a cycle time of 20ms.

\section{CONCLUSIONS}
This paper presents an MPC-based energy management strategy for the air conditioning system of an electrified vehicle. Starting from a validated model of the vapor compression refrigeration system that predicts its nonlinear dynamics and the energy consumption of the actuators, a constrained optimal control problem is formulated with the objective of jointly reducing the energy use while tracking a target evaporator pressure. Dynamic Programming is then applied to compute and analyze the global optimal solution over a prescribed drive cycle and A/C cooling load.

A real-time implementation is then achieved by approximation of the value function obtained from DP, which is imposed as the terminal cost of a receding horizon optimal control problem. This allows for the prediction horizon to be reduced to a single step, leading to a sub-optimal solution that achieves excellent performance on reference tracking and energy use reduction when compared against DP for different drive cycles and thermal load requirements.

The current work focuses on designing an adaptation law such that the tuning of the weight $\alpha'$ in the objective function is performed online. Next, implementation in a test vehicle and experimental verification will be conducted to assess the real-time capabilities of the proposed MPC.

\bibliographystyle{IEEEtran}
\bibliography{AC_system}

\end{document}